\documentclass[%
 reprint, 
 amsmath,amssymb,
 aps, physrev,
floatfix,
]{revtex4-2}

\usepackage{graphicx}
\usepackage{dcolumn}
\usepackage{bm}
\usepackage{xcolor}
\usepackage{amsmath}

\begin{document}

\preprint{APS/123-QED}

\title{\textbf{Transition Waves for Energy Trapping and Harvesting } 
}

\author{Sneha Srikanth}
\author{Andres F. Arrieta}%
 \email{Contact author: aarrieta@purdue.edu}
\affiliation{%
 School of Mechanical Engineering, Purdue University, West Lafayette, Indiana 47907, USA.
}%

\date{\today}

\begin{abstract}
The presence of multiple stable states and associated nonlinear phenomena, such as hysteresis, in multistable mechanical metamaterials enables frequency-independent energy harvesting and shock absorption. This study focuses on shock absorption achieved by locking transition waves to trap energy at designed locations within a multistable metamaterial. We further demonstrate that the same system can simultaneously harvest energy from impact loading, thereby exhibiting multifunctionality. 
The model of the multistable metamaterial is a one-dimensional chain of bistable units whose transition wave dynamics are related to topological solitary waves governed by the $\phi^4$ equation. This connection enables analytical estimation of critical design parameters required for energy trapping and also the amount of energy trapped. Numerical simulations and experiments show that trapping energy in transition waves leads to enhanced damping performance compared to corresponding linear metamaterials. We further propose design variations to increase the amount of energy trapped in the transition wave. Additionally, we identify energy splitting as a damping mechanism that arises when there are repeated impulses or a single high-amplitude impulse that generates multiple transition waves. The transition waves interact to produce localized, fast-dissipating breathers, leading to a damped response. Furthermore, experiments demonstrate that multistable metamaterials can simultaneously achieve improved energy harvesting and better damping performance compared to their linear counterparts. Together, these results highlight the use of transition waves for creating multifunctional multistable metamaterials.
\end{abstract}


\maketitle


\section{\label{sec1}Introduction}

Mechanical metamaterials are artificially structured materials composed of units designed to exhibit unconventional properties, such as negative Poisson's ratio \cite{lakes_foam_1987,silberschmidt_metamaterials_2015} and negative effective mass density \cite{huang_negative_2009}. These unique properties, along with recent advances in additive manufacturing, have paved the way for various applications of metamaterials, including energy harvesting \cite{daqaq_role_2014}, soft robotics \cite{rafsanjani_programming_2019}, noise control \cite{gao_acoustic_2022}, and shock absorption \cite{correa_negative_2015,kim_impact_2017}. 

In general, materials for shock absorption rely either on irreversible processes, such as plasticity in metals and fracture in ceramics and polymers, or have low stiffness, such as foams and gels \cite{yu_chapter_2022}. The former implies degrading, or even non-reusable systems, while the latter severely restricts the energy of the shocks that can be absorbed. Compared to traditional shock absorbing materials, mechanical metamaterials offer several advantages, including predictable and repeatable behavior, and good damping performance with limited impact on their structural stiffness \cite{pechac_metamaterial_2022}. These advantages arise because the damping mechanisms in metamaterials rely on the unit's geometric design rather than the constituent material. Furthermore, metamaterials can be designed so that the damped energy is not dissipated but instead harvested, enabling multifunctional behavior \cite{lu_dual-functional_2021, dwivedi_simultaneous_2020}. 

Different phenomena can be leveraged to design metamaterials for shock absorption or energy harvesting. Bragg scattering \cite{brillouin1946wave} and local resonance \cite{liu_locally_2000} are two phenomena that have been extensively used to block the propagation of certain frequencies through the metamaterial \cite{hussein_metadamping_2013} by creating a band gap. In particular, locally resonant metamaterials can also be designed for enhanced energy harvesting over a targeted range of frequencies \cite{ma2020acoustic}.
However, shock excitations contain a wide range of frequencies that can lie outside of the band gap. Even if multiple metamaterials tuned with different band gaps are utilized, harvesting or damping low frequencies using these two mechanisms requires impractically large units (for scattering) or negligibly small stiffness/ extremely heavy units (for local resonance). 

Another class of metamaterials utilizes negative stiffness elements, such as buckled beams, perforated plates, and shells, for increased energy absorption while maintaining high structural stiffness \cite{lakes2001extreme,tan2024negative}. Negative stiffness elements leverage uniquely nonlinear phenomena, like hysteresis, energy trapping, and snap-through, as frequency-independent mechanisms for shock absorption \cite{frenzel_tailored_2016,ha_design_2018,shan_multistable_2015,yang_multi-stable_2019} and energy harvesting \cite{ramlan2010potential,hwang_input-independent_2018,zhou2022multistable}. They can undergo different loading and unloading paths on the force-displacement diagram and absorb energy. Furthermore, input-independent energy harvesting is achieved when the elements snap through under an input force, converting quasi-static or low-frequency sinusoidal inputs to a high-velocity response accompanied by vibrations at a constant, high frequency \cite{hwang_input-independent_2018, hwang2022topological}. 

Negative stiffness elements can be bistable, i.e., they can possess two stable equilibrium states. Metamaterials composed of bistable units can possess several stable states and are referred to as multistable metamaterials. One pathway to energy trapping in multistable metamaterials involves the transition of the bistable units from a lower to a higher energetic state under an external force \cite{shan_multistable_2015}. A disadvantage of this method is that energy can be trapped only when the force acts in one direction and not the other. 

Another pathway for energy trapping that does not rely on uneven potential wells in multistable metamaterials is to grade the inter-unit stiffness or the onsite bistable potential and thereby modify the potential energy landscape \cite{ramakrishnan2020transition, jin_guided_2020}. The energy trapping mechanism relies on spatially localizing the energy imparted to the metamaterial by stopping the generated transition waves (solitary waves that alter the phase of the metamaterial as they travel \cite{nadkarni2014dynamics}) in a region of local potential energy minimum.
An advantage of energy trapping using transition waves is this method naturally lends itself to wave-based analytical methods \cite{nadkarni2014dynamics} that facilitate the analysis and design of large metamaterials for energy trapping without computationally intensive finite element simulations. 

In the present work, we investigate energy trapping using transition waves in metamaterials comprising one-dimensional chains of bistable units subjected to an impulsive excitation on one end. Using the $\phi^4$ formalism for topological solitary waves
, we analyze effects of the metamaterials' parameters (i.e., mass and stiffness distribution) on the energy carried by the transition waves. Our analysis allows us to establish the conditions for energy trapping at any desired location in the metamaterial using carefully designed defects. Comparison against the response of a corresponding linear metamaterial demonstrates the superior effective damping performance offered by the multistable metamaterial under impulsive loading in simulations and experiments. Under large-amplitude or repeated impulses, the metamaterial exhibits a different damping mechanism, referred to as energy splitting, arising from the interaction between fast-dissipating breathers. Finally, we demonstrate broadband energy harvesting from the transition wave. Our results show that multistable metamaterials enable superior energy absorption capabilities compared to linear counterparts, while offering multifunctionality by leveraging the input-independent energy harvesting afforded by the dynamics of transition waves.   

The rest of the paper is organized as follows: Section~\ref{sec:analytical} presents the model for the 1D multistable metamaterial and derives the conditions for energy trapping. Section~\ref{sec:ET} shows results on energy trapping from numerical simulations while in Sec.~\ref{sec:ES}, we use the model to investigate the phenomenon of energy splitting. Section~\ref{sec:expt} validates our findings on energy trapping in experiments. Section~\ref{sec:mult} demonstrates the multifunctional metamaterial capable of simultaneous damping and energy harvesting. We finally conclude our study in Sec.~\ref{sec:conc}.

\section{\label{sec:analytical}Theoretical Analysis}
\begin{figure}
\includegraphics[width=\linewidth]{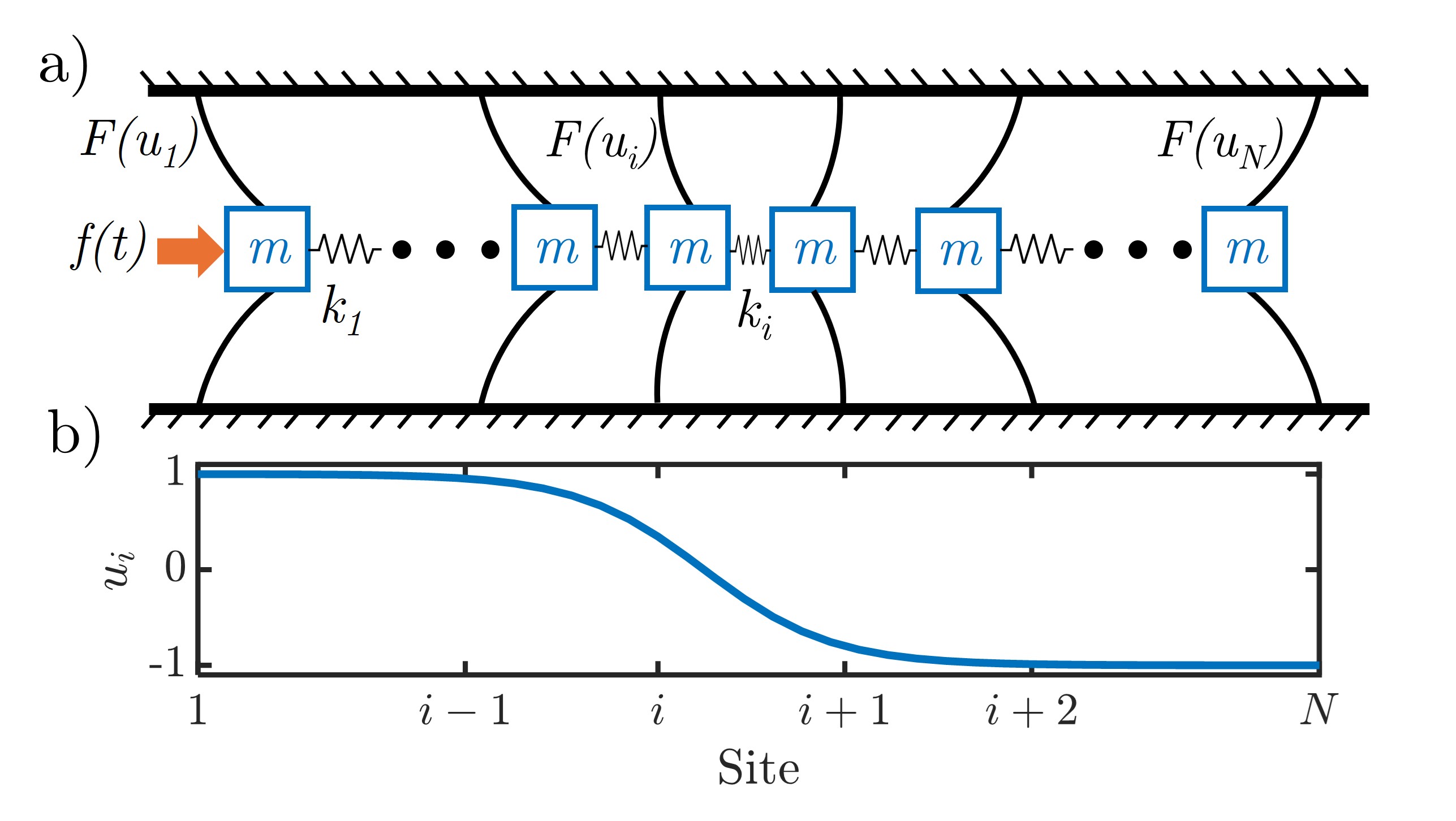}
\caption{\label{fig:schematic} a) Schematic representation of the multistable metamaterial composed of masses connected by intersite springs, with each mass also connected to a bistable substrate potential. A force on the first unit can trigger a transition wave. b) Displacement profile of a transition wave propagating through the structure.}
\end{figure}

We model the multistable metamaterial as a chain of $N$ units (Fig.~\ref{fig:schematic}a), each connected to a bistable substrate or on-site potential. The governing equation for the displacement $u_i$ of the $i^\text{th}$ unit in the metamaterial is given by 
\begin{align}
    m_i\ddot{u}_i+b_i\dot{u}_i+k_{i-1}(u_i-u_{i-1})+k_i(u_i-u_{i+1}) \nonumber\\ +F_i(u_i) =0.   
    \label{eq:main}
\end{align}

Here, $m_i$ and $b_i$ are the mass and the damping coefficient of the $i^\text{th}$ site. The onsite force due to the bistable potential at the $i^\text{th}$ site is $F\left(u_i\right)=B_iu_i\left(u_i^2-d_i^2\right)$, whose strength is determined by the constant $B_i$. The bistable potential's stable equilibria are $\pm d_i$. Any two consecutive units, for example at sites $i$ and $i+1$, are connected by intersite elastic springs of stiffness $k_i$.
A transient force (shown as $f(t)$ in Fig.~\ref{fig:schematic}a) or an initial velocity imparted to the first unit can trigger the propagation of a transition wave (Fig.~\ref{fig:schematic}b). 

Energy trapping occurs when a transition wave is trapped at a site, storing the input energy as strain energy \cite{jin_guided_2020}. This analysis aims to determine the condition required for energy trapping. To simplify our analysis, we consider all units to have the same properties with the exception of the intersite stiffnesses, which we use to modify the potential energy landscape of the metamaterial. Non-dimensionalizing the governing equation~\eqref{eq:main} under these assumptions yields
\begin{align}    \bar{u}_{i,\bar{t}\bar{t}}+\bar{b}\bar{u}_{i,\bar{t}}+(2r_i\bar{u}_i-r_{i-1}\bar{u}_{i-1}-r_{i+1}\bar{u}_{i+1})\nonumber\\+\omega_0^2\bar{u}_i(\bar{u}_i^2-1) =0, \label{eq:main_nondim}
\end{align}
where $\bar{u}_i=u_i/d$, $\bar{t}=t\sqrt{k_1/m}$. The normalized damping coefficient is $\bar{b} = \tfrac{b}{\sqrt{k_1m}}$. The intersite stiffness is normalized to a stiffness ratio of $r_i=k_i/k_1$. Finally, $\omega_0 = \sqrt{\tfrac{Bd^2}{k_1}}$ is a design parameter indicating the discreteness of the metamaterial \cite{hwang_solitary_2018}. 

We begin our analysis starting from the equation for a uniform multistable metamaterial ($r_i=1$):
\begin{align}    \bar{u}_{i,\bar{t}\bar{t}}+\bar{b}\bar{u}_{i,\bar{t}}+(2\bar{u}_i-\bar{u}_{i-1} &-\bar{u}_{i+1})\nonumber\\&+\omega_0^2\bar{u}_i(\bar{u}_i^2-1) =0. \label{eq:main_uniform}
\end{align}

Let $a$ be the unit size. Taking continuum limit (with the nondimensional position $\bar{x} =x/a$ of the $i^\textrm{th}$ site being $\bar{x}=i$),
\begin{align}
    \bar{u}_{\bar{t}\bar{t}}+\bar{b} \bar{u}_{\bar{t}}-\bar{u}_{\bar{x}\bar{x}}+\omega_0^2\bar{u}(\bar{u}^2-1) =0.\label{eq:cont}
\end{align}
The equation can be further scaled to
\begin{align}
    U_{TT}+\gamma U_{T}-U_{XX}+U(U^2-1) =0, \label{eq:phi4}
\end{align}
where $X = \omega_0\bar{x}$, $U = \bar{u}$, and $T =\omega_0\bar{t}$.
In the absence of material damping, Eq.~\eqref{eq:phi4} has a transition wave solution, also known as a kink, given by
\begin{align}
    U_K = \pm \tanh{\left(\frac{X-X_0}{\sqrt{2(1-V^2)}}\right)}, \label{eq:TW}
\end{align}
where $X_0$ is the normalized position and $V$ is the normalized speed of the transition wave \cite{lizunova2021introduction}. Depending on the sign, the transition wave is either rarefactive (+) or compressive (-). They are also referred to as kink or antikink, respectively. In dimensional values, the transition wave’s displacement profile is
\begin{align}
    u_K = \pm d \tanh{\left(\frac{\omega_0(x-x_0)}{a\sqrt{2(1-(\tfrac{v}{v_m})^2)}}\right)},
\end{align}
which shows that $v_m = a\sqrt{k/m}$ is the theoretical maximum speed of the transition wave, whose corresponding nondimensional value is $V_m=1$. A better estimate of the transition wave's maximum speed when triggered by an impulse is given by the speed with which perturbations travel through the metamaterial. Linear waves carry energy through the metamaterial with speed $c_{g,\textrm{max}}$, the maximum group velocity, given by 
\begin{align}
    c_{g,\textrm{max}} = v_m\sqrt{\omega_0^2+1-\omega_0\sqrt{\omega_0^2+2}}. \label{eq:cg}
\end{align}
Algebraic manipulation of the above expression shows that $c_{g,\textrm{max}}\in\left[0,v_m \right]\ \forall\ \omega_0\geq0$. The speed of the transition wave approaches the theoretical maximum of $v_m$ only as $\omega_0\rightarrow0$, i.e., when the strength of the onsite potential is extremely weak (or when the intersite stiffness is extremely high). 

The energy carried by the transition wave can be estimated by integrating Eq.~\eqref{eq:phi4} with the solution Eq.~\eqref{eq:TW}, yielding the nondimensionalized energy as
\begin{equation}
   \bar{E} =\frac{2\sqrt{2}}{3a^2 \sqrt{1-V^2 }} d^2 \omega_0. \label{eq:Enondim}
\end{equation}
In dimensional values, this is 
\begin{equation}
   E =\frac{2\sqrt{2}}{3\sqrt{1-\tfrac{mv^2}{ka^2} }} d^3 \sqrt{Bk}. 
\end{equation}
The maximum amount of energy is carried by the transition wave when its speed is maximum ($v=c_{g,\textrm{max}}$), that is when 
\begin{align}
    E_{\textrm{max}} &=\frac{2\sqrt2}{3\sqrt{\omega_0\sqrt{\omega_0^2+2}-\omega_0^2}}d^3\sqrt{Bk}, \label{eq:Emax}
\end{align}
whereas the energy is minimum when the transition wave is stationary. This value is also the amount of energy that can be trapped as strain energy in the metamaterial, given by
\begin{equation}
    E_{\textrm{min}}=\frac{2\sqrt2}{3}d^3\sqrt{Bk}.  \label{eq:Emin}
\end{equation}


Reducing the intersite stiffness in a section of the metamaterial, for example by including a region of softer intersite springs of stiffness $k_s<k$, creates a local minimum for the energy carried by the transition wave. Given sufficient difference between $k$ and $k_s$, a transition wave entering the softer region can lose energy to such an extent that it can no longer reenter the relatively stiffer region. In this way, we can trap the transition wave in a desired location. We can derive the condition for energy trapping by taking the transition wave's maximum energy in the softer section to be less than its minimum energy in the stiffer section, i.e., $E_\textrm{max,soft} <E_\textrm{min,stiff}$. Using Eq.~\eqref{eq:Emax} and Eq.~\eqref{eq:Emin} yields
\begin{align}
    \frac{2\sqrt{2}}{3\sqrt{\sqrt{\frac{Bd^2}{k_s}}\sqrt{\frac{Bd^2}{k_s}+2}-\frac{Bd^2}{k_s}}}d^3\sqrt{Bk_s} < \frac{2\sqrt{2}}{3}d^3\sqrt{Bk}.
\end{align}
Let $r$ be the stiffness ratio $\frac{k_s}{k}$. Then, the above equation can be simplified to 
\begin{align}
    r^3+2\omega_0^2r-2\omega_0^2<0, \label{eq:cubic}
\end{align}
where $\omega_0=\sqrt{Bd^2/k}$ is the discreteness parameter in the stiffer sections of the metamaterial. Equation~\eqref{eq:cubic} gives the condition for energy trapping. The depressed cubic polynomial on the left-hand side of the equation has only one real root, $r^{*}$, which is 
\begin{align}
    r^\ast=\sqrt[3]{\omega_0^2+\sqrt{\omega_0^4+\frac{8\omega_0^6}{27}}}-\sqrt[3]{-\omega_0^2+\sqrt{\omega_0^4+\frac{8\omega_0^6}{27}}}. \label{eq:rcr}
\end{align}

This analysis allows us to establish that energy trapping occurs when $r<r^{\ast}$. The critical value is determined solely by the nondimensional discreteness parameter $\omega_0$, whose inverse-square $1/\omega_0^2 = k/Bd^2$ can be interpreted as the intersite stiffness of the metamaterial normalized by the onsite stiffness. Eq.~\eqref{eq:rcr} is verified against numerically obtained values of $r^\ast$ in the following section. Considering material damping or added damping (e.g., adding on-site dashpots to some units in the metamaterial) would further slow down the transition wave, reducing $E_\textrm{max,soft}$ and further increasing $r^*$. 


\section{\label{sec:numerical}Numerical simulations}
The preceding analysis provides us with a tractable analytical model for exploring the dynamics of our metamaterial with a particular focus on elucidating its capacity for energy trapping and reduced transmissibility. We now numerically investigate energy trapping in the model by including regions of soft defects. We integrate the governing equations \eqref{eq:main_nondim} on MATLAB with ode45 using a relative tolerance of $10^{-8}$ and absolute tolerance of $10^{-9}$. Numerical simulations are performed with the discreteness parameter set to $\omega_0 = \sqrt{0.1}$ and ignoring material damping ($\bar b=0$) unless otherwise specified. Free boundary conditions are applied at both ends of the metamaterial containing $N=500$ units. Impulsive forcing is implemented by giving the first unit an initial velocity of $v_0$ (or nondimensional value $\bar v_0 = v_0\sqrt{m/kd^2}$). Section I of the Supplemental Material shows that an initial velocity generates a transition wave when it is above a critical value, which is approximately 1.15 for $\omega_0=\sqrt{0.1}$. We consider a value of initial velocity of $\bar v_0=2$ in the simulations, which is sufficiently high to generate a transition wave. 

\begin{figure*}
    \centering
    \includegraphics[width=\linewidth]{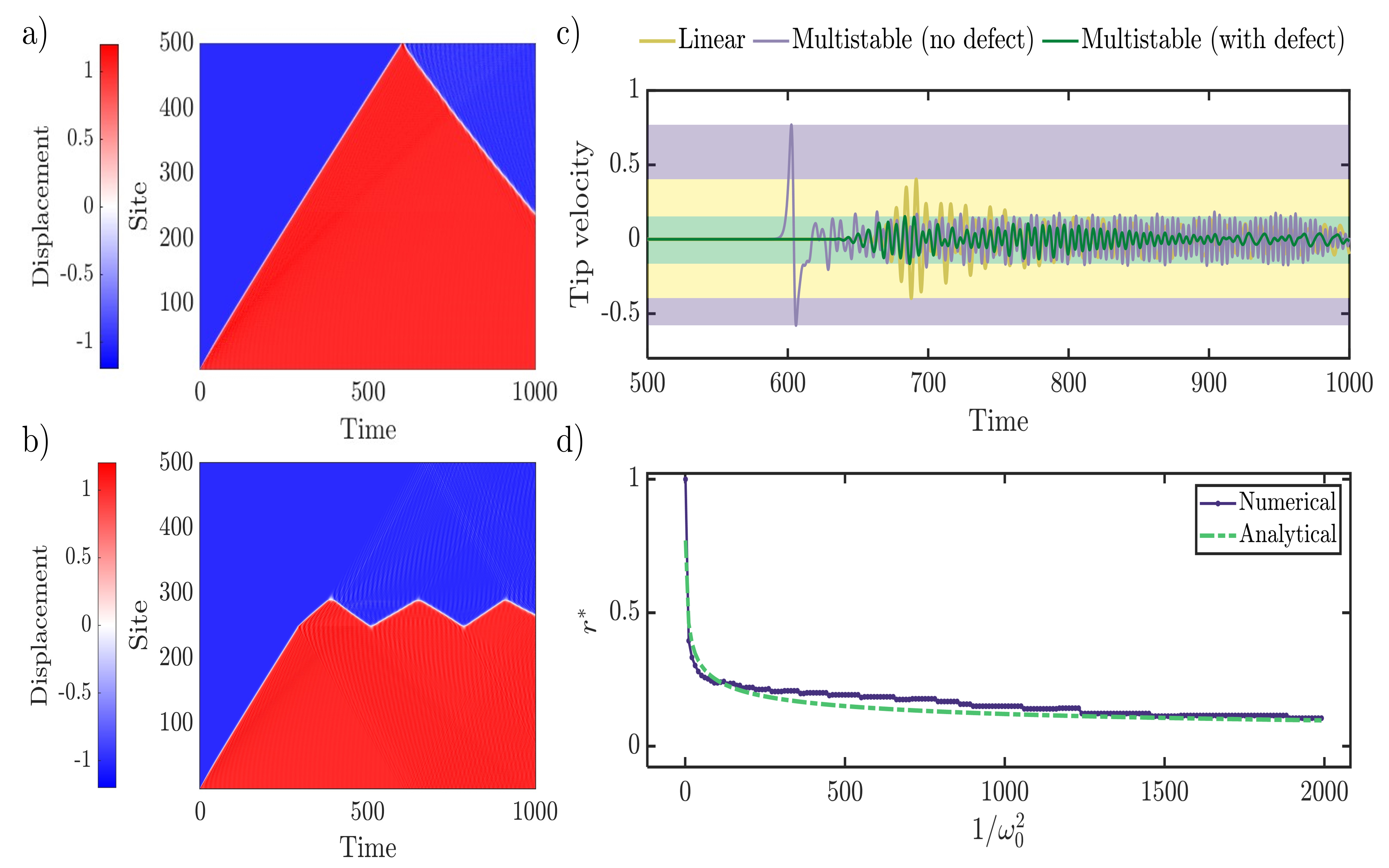}
    \caption{a) Transition wave propagation in the absence of defect. b) Energy trapping achieved by defect of $r=0.35$ from sites 250 to 290. c) Reduced tip velocity due to energy trapping in multistable metamaterial with defect compared to linear metamaterial (with defect) and multistable metamaterial without defect. d) Critical stiffness ratio ($r^*$) for energy trapping as a function of the normalized intersite stiffness ($1/\omega_0^2$).}
    \label{fig:ET}
\end{figure*}

\subsection{Energy trapping using soft defects in intersite stiffness}\label{sec:ET}

\subsubsection{Single region of defect}
A uniform metamaterial without defects allows transition waves to propagate through the metamaterial (Fig.~\ref{fig:ET}a). Including a defect in the form of reduced intersite stiffness in any desired region within the metamaterial traps the transition wave. For example, the transition wave is trapped between sites 250 and 290 by including a stiffness defect of $r=0.35$ at these sites, as seen in Fig.~\ref{fig:ET}b. Interacting with defects causes the transition wave to lose some energy through radiation of phonons, which is the emission of small amplitude linear waves \cite{malomed1992perturbative}. We can observe the emission of these linear waves as the transition wave reflects back and forth within the soft defect region (Fig.~\ref{fig:ET}b). Radiation loss further promotes energy trapping. 

As a result of trapping the transition wave, less energy is transmitted to the last unit of the metamaterial and its velocity (denoted as tip velocity in Fig.~\ref{fig:ET}c) is significantly reduced. The multistable metamaterial exhibits an 80 \% reduction in tip velocity due to energy trapping in the carefully designed defect region. We further compare its behavior with that of a corresponding linear metamaterial of similar properties (including the defect region), with the only difference being that the onsite force is linearized about the equilibrium $-d_i$ to $F_i(u_i) = 2B_id_i^2(u_i+d_i)$ in Eq.~\eqref{eq:main} and $2\omega_0^2(\bar u_i+1)$ in Eq.~\eqref{eq:main_nondim}. The linear metamaterial serves as an additional benchmark to evaluate the damping performance of the multistable metamaterial due to energy trapping. This comparison shows that the multistable metamaterial with defect has a tip velocity of 37\% of that of the corresponding linear metamaterial (Fig.~\ref{fig:ET}c). This reduction is explained by the reliance of the linear metamaterial on dispersion to reduce the maximum tip velocity. On the other hand, the multistable metamaterial with defect utilizes the more effective trapping of the transition wave as well as dispersion of the trailing phonons to achieve tip velocity reduction.  

Energy trapping occurs when the stiffness ratio $r$ is less than a critical value $r^*$ (as discussed in Sec.~\ref{sec:analytical}). Figure \ref{fig:ET}d illustrates how the critical stiffness ratio in an undamped multistable metamaterial varies with the normalized intersite stiffness $1/\omega_0^2$. The critical stiffness ratio reduces as the intersite stiffness increases. This is because, as the metamaterial becomes `smoother' (i.e., less discrete), the transition wave's kinetic energy increases. Consequently, the defect region must be significantly softer to compensate for this energy increase for trapping the transition wave. The critical stiffness ratios determined by Eq.~\eqref{eq:rcr} agree well with the values obtained from numerical simulations, especially for $1/\omega_0^2 \gg 1$. Our analysis allows us to predict the behavior of a metamaterial of several units and design defect regions for trapping transition waves.    

\begin{figure*}
    \centering
    \includegraphics[width=\linewidth]{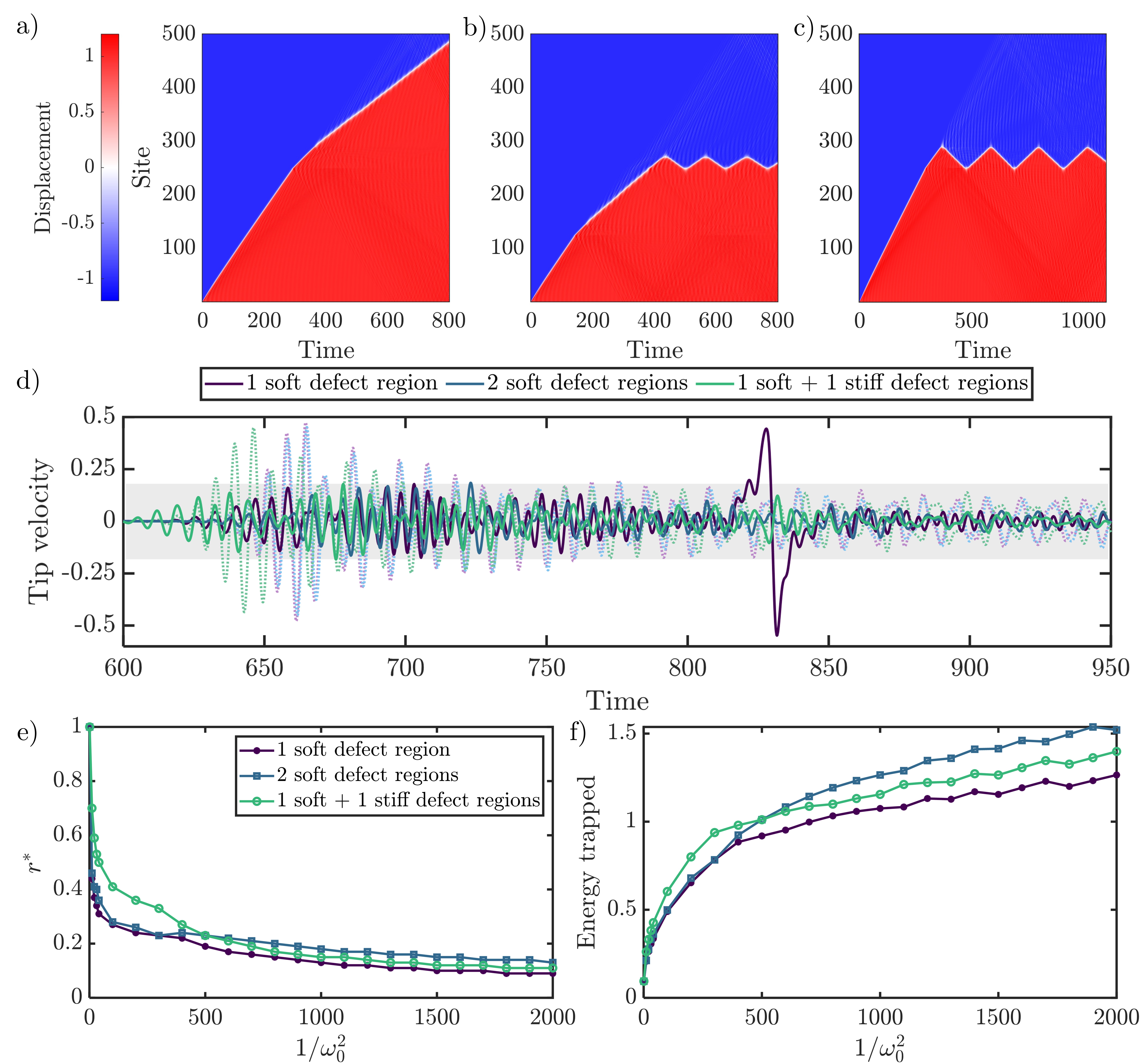}
    \caption{(a) Propagation of transition wave in the presence of a defect region from sites 250 to 290. b) Energy trapping using two soft defect regions, at sites 125-145 and 250-270. c) Energy trapping using a soft defect region at sites 250-290 followed by a stiff defect region of $r=2$ at sites 290-330. All soft defect regions have $r=0.5$. d) Tip velocities of the multistable metamaterial for the three cases shown in a-c (solid lines) and those of the corresponding linear metamaterial (dotted lines). Grey shaded area highlights the range of tip velocities for cases in b and c. e) Critical stiffness ratio and f) amount of strain energy trapped in the transition wave as a function of the normalized intersite stiffness $1/\omega_0^2$ for the three types of defect regions shown in a-c. $1/\omega_0^2=10$ in a-c.}
    \label{fig:improvingET}
\end{figure*}

\subsubsection{Methods to improve energy trapping}\label{sec:improvingET}

A single region of defect can require a considerably small stiffness ratio ($r\ll 1$) to trap the transition wave, especially when we would like the metamaterial to have high intersite stiffness, as indicated by the asymptotic behavior of $r^{*}$ as $1/\omega_0$ tends to infinity (Fig.~\ref{fig:ET}d). We can also infer the decrease in $r$ by simplifying Eq.~\eqref{eq:rcr} with $\omega_0 \to 0$. This yields $r^\ast \approx (2\omega_0^2)^{1/3}$, which decreases quickly to zero as the normalized intersite stiffness approaches infinity, i.e., $1/\omega_0^2 \to \infty$. This reduction in $r$ is disadvantageous since the amount of energy trapped, which is proportional to $\sqrt{k_s}$ (as indicated by Eq.~\eqref{eq:Emin}), decreases as the defect region becomes softer. Small intersite stiffness also makes the multistable metamaterial less stiff and more prone to fracture in practical applications. 

One method to improve the critical stiffness ratio is to split the defect region into multiple parts, which causes the transition wave to repeatedly interact with defects. Every interaction causes the transition wave to lose energy to radiation. For example, a stiffness ratio of 0.5 is not sufficient for energy trapping in a multistable metamaterial of $1/\omega_0^2 = 10$ with a single region of defect (Fig.~\ref{fig:improvingET}a). Including two separate defect regions enables trapping of a transition wave with the same initial energy (Fig.~\ref{fig:improvingET}b). 

A second method to improve the critical stiffness ratio is to include a stiff defect at the end of a soft defect region. This method works by increasing the energy barrier for the transition wave to exit the soft defect region. Figure~\ref{fig:improvingET}c shows energy trapping achieved with a region of soft defect followed by a region of stiff defect. Stiff defects have been used to reflect transition waves within a 1D multistable metamaterial \cite{hwang_solitary_2018} and to redirect and pin them in a 2D metamaterial \cite{jin_guided_2020}. Here, soft defects and stiff defects are utilized together to trap a transition wave in a 1D metamaterial, thereby increasing its effective energy dissipation capacity.  

Trapping of transition waves has a direct effect on the reduction in tip velocity of the metamaterial. Figure~\ref{fig:improvingET}d shows the tip velocity of the multistable metamaterial for all the three cases depicted in Figs.~\ref{fig:improvingET}a-\ref{fig:improvingET}c using solid lines. Lighter dotted lines denote the tip velocities in the corresponding linear metamaterial. The maximum tip velocity does not vary significantly between the three cases in the linear metamaterial. The propagation of the transition wave until the end due to a soft defect region of insufficiently low stiffness ratio causes a higher tip velocity in the multistable metamaterial, compared to its linear counterpart. Conversely, energy trapping through the inclusion of two separate soft defect regions, or a soft defect region followed by a stiff defect region, decreases the tip velocity by 59\% compared to their linear counterparts. The smaller range of tip velocity due to these two methods is highlighted by a gray shaded region in Fig.~\ref{fig:improvingET}d.

Figure~\ref{fig:improvingET}e shows the critical stiffness ratios as a function of the normalized intersite stiffness ($1/\omega_0^2$) for three different cases: when a single region of soft defect is present in the metamaterial, when there are two separate regions of soft defect, and when a region of soft defect is followed by a region of stiff defect. The latter two cases result in higher values of critical stiffness ratios. These correspond to an improvement in the amount of energy trapped in the metamaterial (Fig.~\ref{fig:improvingET}f), which depends on the stiffness ratio of the stiff defect or the number of soft defect regions used, and the normalized intersite stiffness of the metamaterial. In Fig.~\ref{fig:improvingET}f, including a stiff defect region after the soft defect region is a better method to improve the amount of energy trapped when the metamaterial is relatively discrete (i.e., when $1/\omega_0^2$ is small). On the other hand, when the metamaterial is more continuous, including two separate defect regions has a larger improvement in the energy trapped (about 25\% as compared to about 10\%).  




\begin{figure}
    \centering
    \includegraphics[width=\linewidth]{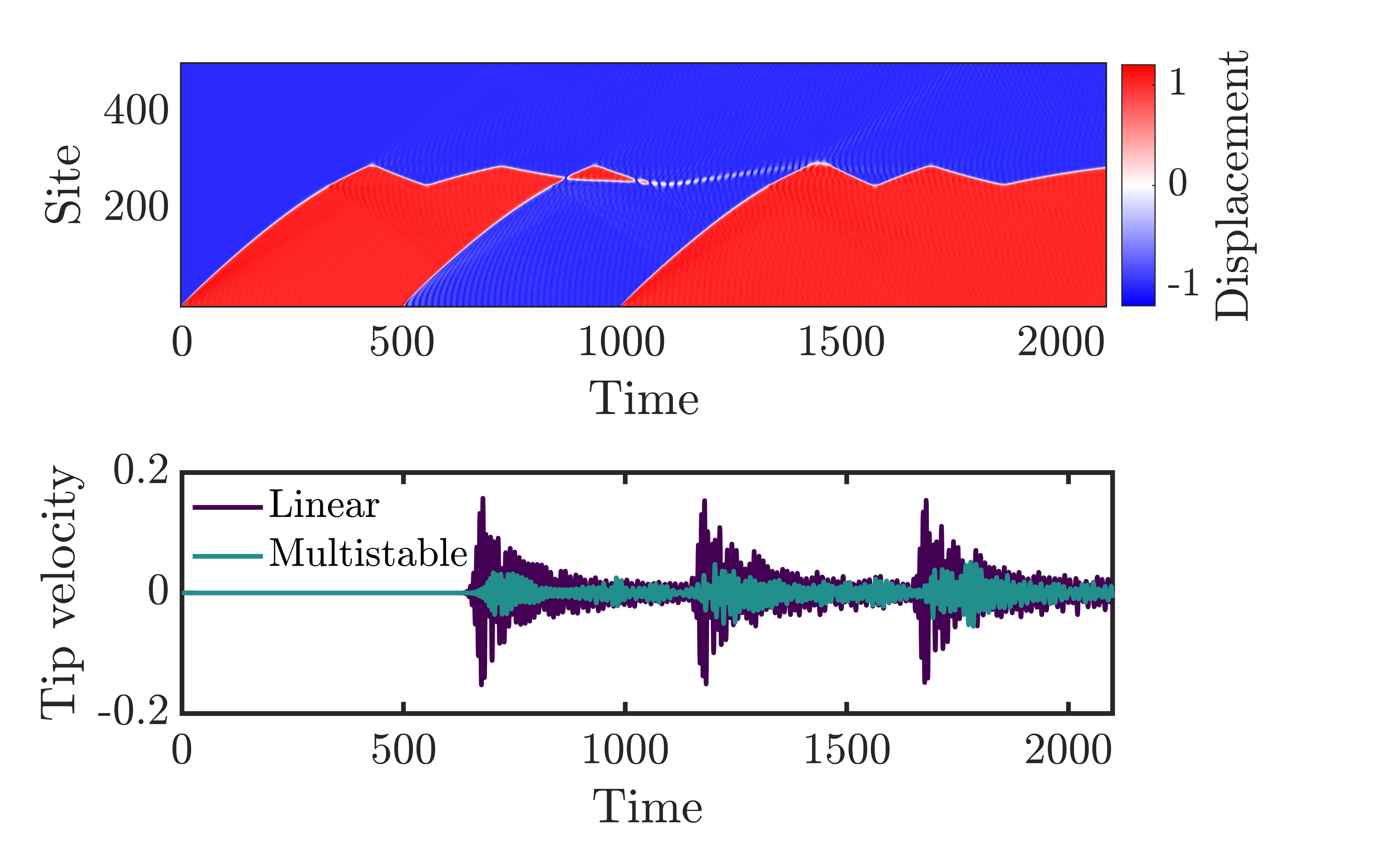}
    \caption{a) Energy trapping and splitting due to three separate impulses (at times 0, 500, and 1000). b) Tip velocity in the multistable metamaterial compared to the corresponding linear metamaterial. $r=0.4$, $\bar b = 0.003$, $\bar v_0=2$.}
    \label{fig:multipleimp}
\end{figure}

\begin{figure*}
    \centering
    \includegraphics[width=\linewidth]{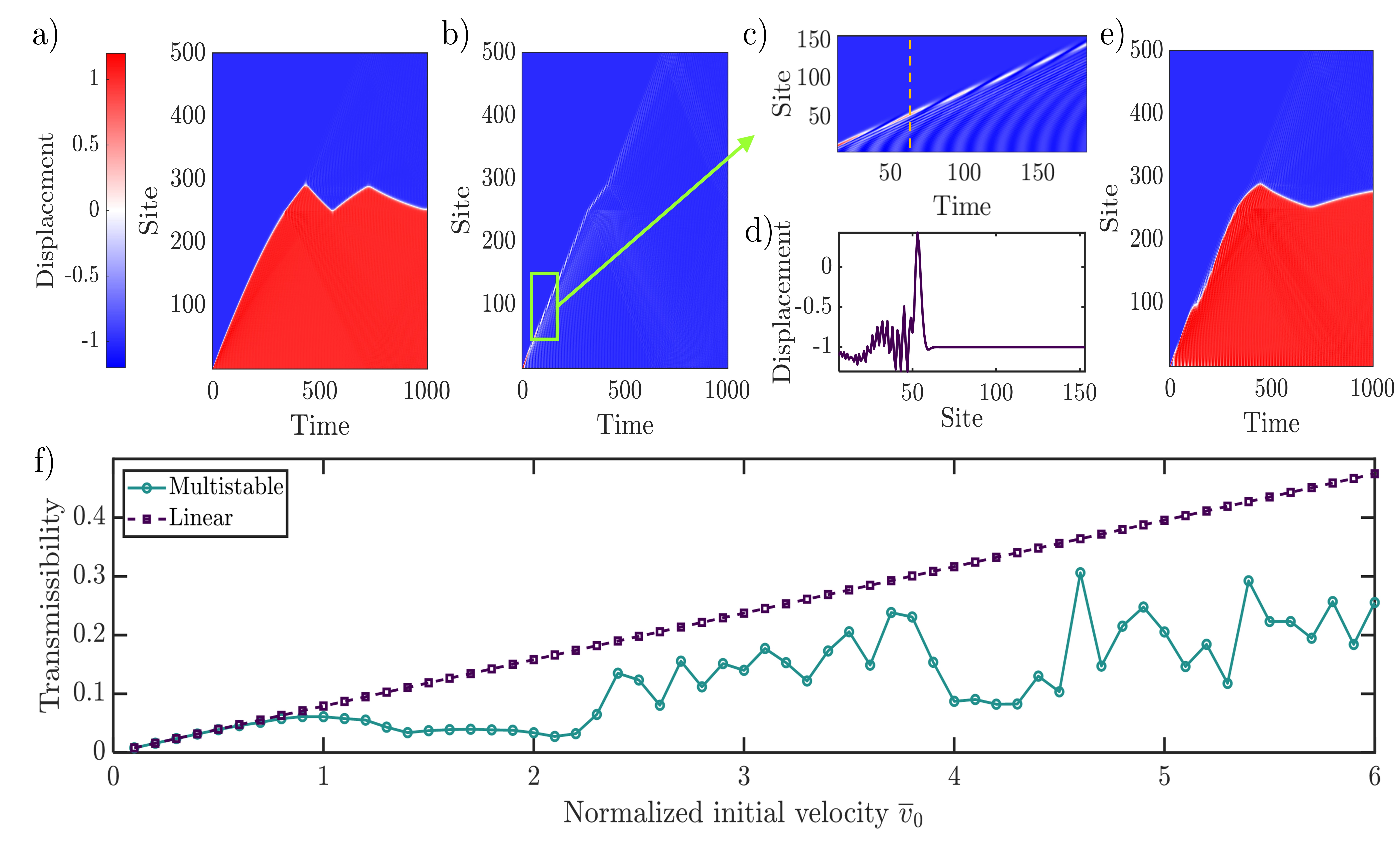}
    \caption{a) Transition wave generation and trapping for $\bar v_0=2$. b) Energy splitting due to generation of two transition waves that combine into a moving breather for $\bar v_0=3$. c) Zoomed outset showing the moving breather in (b). d) Displacement profile of the breather at time $\bar t = 70$ (marked by dashed line in c). e) Energy splitting and energy trapping due to generation of three transition waves for $\bar v_0=4$. f) Transmissibility of the multistable and the corresponding linear metamaterial as a function of the normalized initial velocity $\bar v_0$ imparted by the impulse. $r=0.4$, $\bar b = 0.003$ in all plots.}
    \label{fig:highamp}
\end{figure*}

\subsection{Energy splitting: Behavior under high-amplitude or repeated impulses} \label{sec:ES}

Studies on kink-antikink collisions in the $\phi^4$ system show that a kink and an antikink form a breather on collision \cite{goodman2005kink,campbell1983resonance}. This breather is a large-amplitude unstable localized wave that dissipates over time by radiating phonons. Adding material damping further aids its dissipation. We can leverage this dynamical behavior to damp the response to high-amplitude or repeated impulses in our multistable metamaterial. 

Consider the case of repeated impulses (Fig.~\ref{fig:multipleimp}a). Suppose a compressive transition wave (antikink) generated by an impulse is trapped in the defect region of the metamaterial. A second impulse then generates a kink, i.e., a rarefactive transition wave. When the two transition waves collide in the defect region, they generate a breather that is still trapped and dissipates, thereby completely resetting the metamaterial. Each oscillation of the breather dissipates energy by radiating phonons. Since the energy is split into multiple phonon emissions, this dissipation mechanism is referred to as energy splitting. Any further impulse follows this cycle of energy trapping and energy splitting. In a linear metamaterial, energy from the impulse is not split and propagates as a single stream of phonons. As a result, the tip velocity is significantly lower in the multistable metamaterial compared to its linear counterpart. In Fig.~\ref{fig:multipleimp}b, the maximum tip velocity of the multistable metamaterial subjected to three impulse excitations is nearly one-third of that of the corresponding linear metamaterial under identical conditions.

Figure~\ref{fig:highamp} illustrates another type of energy splitting that occurs at high amplitudes of impulsive forcing. When the energy in the impulse exceeds the amount required to generate a transition wave of maximum energy, multiple transition waves are generated that are alternately compressive and rarefactive. Each successive pair of compressive and rarefactive waves collide to form a moving breather. For example, an initial velocity of $\bar v_0=2$ imparted to the first unit is sufficient to generate a transition wave that is trapped in the defect region in the metamaterial (Fig.~\ref{fig:highamp}a). A larger initial velocity of $\bar v_0=3$ generates a compressive and a rarefactive transition wave, which combine into a moving breather, as seen in Fig.~\ref{fig:highamp}b and its zoomed outset Fig.~\ref{fig:highamp}c. The displacement profile of the breather is shown in Fig.~\ref{fig:highamp}d. An even greater initial velocity generates additional transition waves that pair up to form breathers. Any unpaired transition wave is subsequently trapped in the defect region (as shown in Fig.~\ref{fig:highamp}e for $\bar v_0=4$). In this way, energy from a high-amplitude impulse is split into individual breathers or trapped within the defect region. 

We can quantify the damping performance of the multistable metamaterial by measuring its transmissibility, which is the ratio of the maximum magnitude of the tip velocity to the magnitude of the initial velocity imparted to the first unit. Figure~\ref{fig:highamp}f shows the transmissibility of the multistable metamaterial and also the corresponding linear metamaterial for different values of normalized initial velocity. When $\bar v_0<\bar v_{cr}$ ($\bar v_{cr}\approx 1.15$ for $\omega_0^2=0.1$), no transition wave is generated and both the metamaterials have similar levels of transmissibility. Once the initial velocity is sufficient to trigger a transition wave, energy trapping occurs and the multistable metamaterial performs better than the linear metamaterial. Higher values of initial velocity generate multiple transition waves leading to energy splitting. Furthermore, depending on the initial velocity, energy trapping occurs if there is an unpaired transition wave, resulting in dips in the transmissibility of the multistable metamaterial (see around $\bar v_0=2$ and $\bar v_0=4$ in Fig.~\ref{fig:highamp}e). Due to energy splitting, the maximum tip velocity in a multistable metamaterial does not proportionally grow with the initial velocity, in contrast to the linear metamaterial. Thus, energy splitting offers a pathway to reduce transmissibility of high-amplitude impulsive forces in multistable metamaterials.

\section{Experimental Validation of Energy Trapping}\label{sec:expt}
\begin{figure*}
    \centering
    \includegraphics[width=\linewidth]{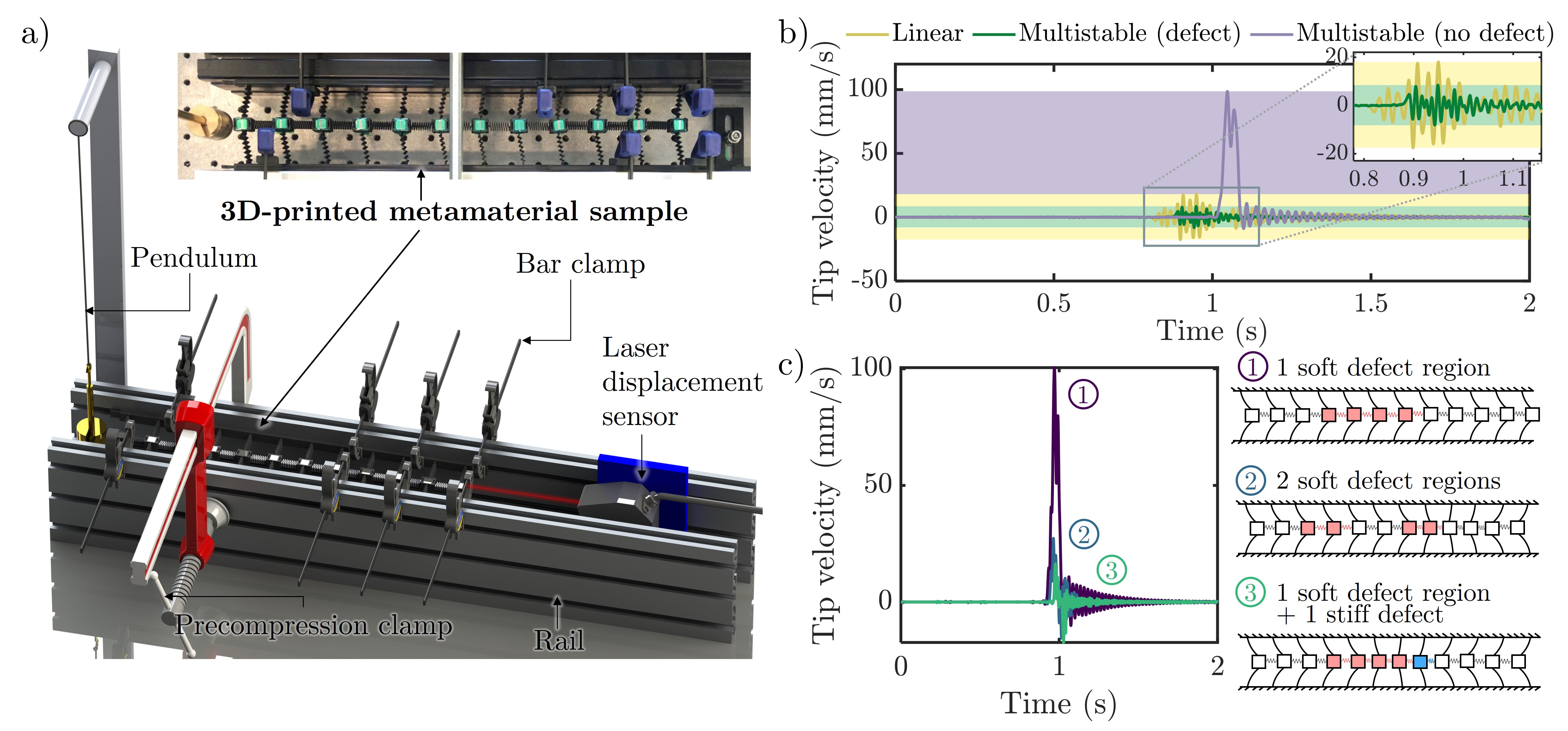}
    \caption{a) Experimental setup with inset showing the 3D-printed metamaterial demonstrator. b) Tip velocities in the linear metamaterial (with a soft defect region of $r=0.4$) and the multistable metamaterial (with and without a soft defect region of $r=0.4$). c) Tip velocity when the multistable metamaterial has a single soft defect region of $r=0.6$ (labeled 1), two separate soft defect regions of $r=0.6$ (labeled 2), and a region of soft defect with $r=0.6$ followed by a stiff defect with $r=1.4$ (labeled 3). The softer (stiffer) intersite springs and the units to their left are colored in red (blue) in the schematic diagrams for the three cases.}
    \label{fig:expt} 
\end{figure*}

We investigate the response of an experimental demonstrator to validate energy trapping. Figure~\ref{fig:expt}a shows the experimental setup. The demonstrator is a multistable metamaterial (shown in inset of Fig.~\ref{fig:expt}a) prepared with a fused deposition modeling 3D printer (Ultimaker S5) using polylactic acid (PLA) filament. The metamaterial contains 12 units. Each unit has a mass weighing 5 grams. The units have slots for inserting intersite springs (Fig.~2 in the Supplemental Material). The height of the intersite spring directly controls the intersite stiffness (Fig.~3 in the Supplemental Material). The intersite springs have a height of 5 mm, unless otherwise specified. 

The bistability of the units arises from precompression of the metamaterial between two rails clamped to an optical table. A pendulum of mass 120 g hung from a mount with a nylon wire of length 45 cm serves as the impactor. The impactor is released from an angle of 20 degrees to excite the metamaterial. The response of the last unit of the metamaterial is measured using laser displacement sensor (Keyence LK-H157). The corresponding linear metamaterial's response is obtained by measuring the tip response in the absence of precompression. 

We obtain the tip velocities of the metamaterial under different conditions (Fig.~\ref{fig:expt}b). When the metamaterial features no intersite stiffness defect, the transition wave propagates all the way through, resulting in a large tip velocity, highlighted by the purple shaded region. In contrast, when a soft defect region of stiffness ratio $r\approx0.4$ is implemented using intersite springs of height 2 mm from sites 4 to 7,  energy trapping results in a 92~\% decrease in the tip velocity (highlighted by the green shaded region). Finally, comparing the response of the multistable metamaterial with that of the corresponding linear metamaterial featuring the same soft defect region (zoomed inset of Fig.~\ref{fig:expt}b), we observe a 50\% reduction in tip velocity relative to the linear metamaterial. 

Figure~\ref{fig:expt}c examines the two methods discussed in Sec.~\ref{sec:improvingET} for improving energy trapping: inclusion of multiple soft defect regions, and incorporating stiff defects after a soft defect region. These methods are compared against the case when a single soft defect region is present in the metamaterial. A soft defect region of $r\approx0.6$ is insufficient to cause energy trapping, leading to a large tip velocity. By including two separate soft defect regions, or by including a stiff defect of $r\approx1.4$ (i.e., an intersite spring of height 7 mm) at the end of a single soft defect region, the tip velocity is reduced in the multistable metamaterial due to trapping of transition waves. Thus, our experiments validate that multistable metamaterials can absorb energy from impact by trapping transition waves in suitably designed defect regions.

\section{Multifunctionality of the Multistable Metamaterial}\label{sec:mult}

We further utilize the metamaterial's response to harvest energy by leveraging the high snap-through speed of the propagating transition wave. This results in a multifunctional metamaterial capable of energy harvesting as well as shock absorption. 

\begin{figure}
    \centering
    \includegraphics[width=\linewidth]{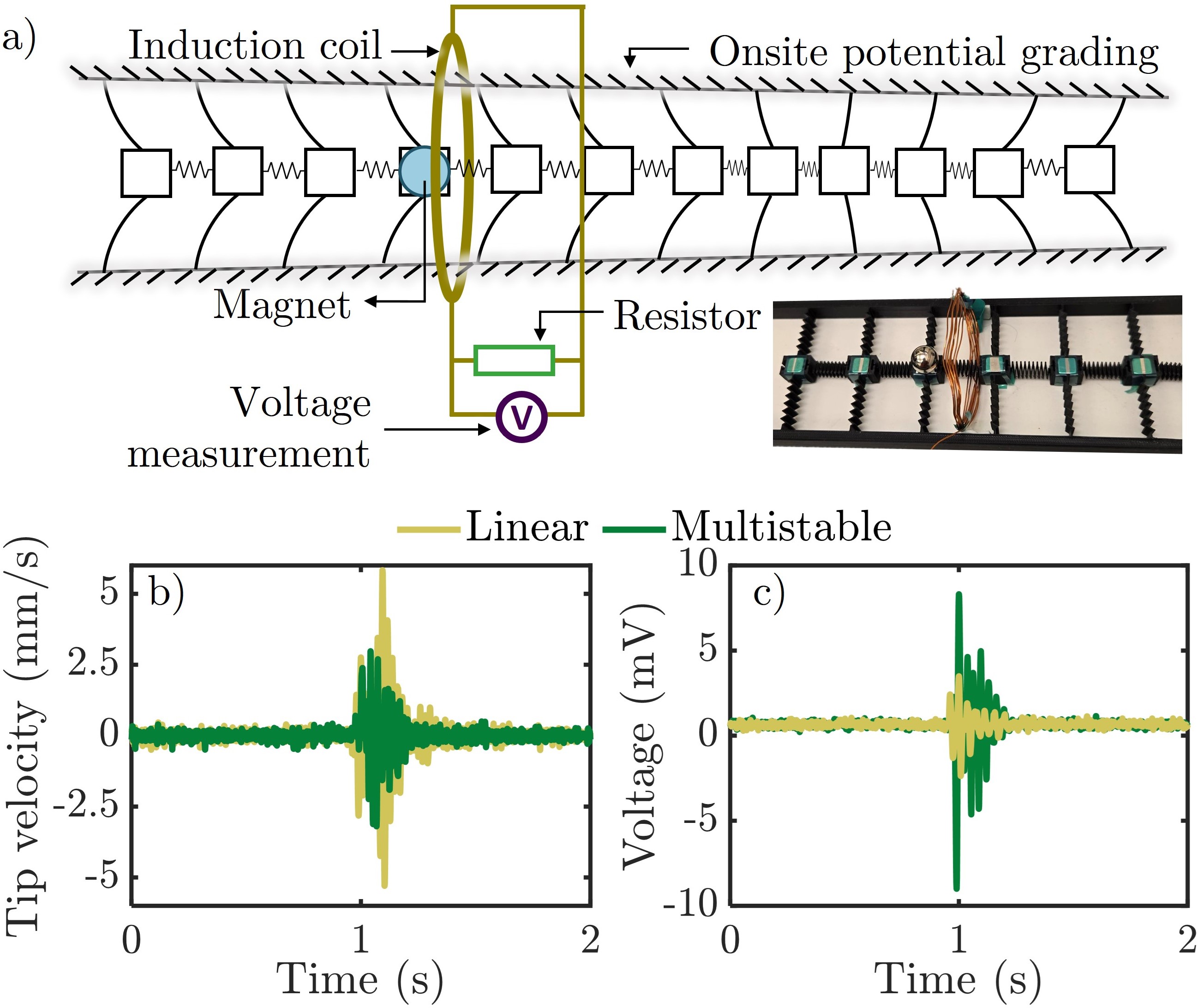}
    \caption{a) Experimental setup for energy harvesting. b) Tip velocities and c) voltages of harvested energy achieved in the multistable and the linear metamaterials.}
    \label{fig:multifunc} 
\end{figure}

We design the metamaterial to reset automatically after harvesting energy from the impact. This can be achieved by altering the potential energy landscape of the metamaterial through spatial variation in the intersite or the onsite stiffness. A stiffening gradation (i.e., a monotonic increase in stiffness) leads to a potential energy minimum at the start of the metamaterial. As a result, any transition wave reflects back to the start instead of propagating forward. This phenomenon is referred to as boomerang effect. We demonstrate the multifunctionality of the multistable metamaterial with boomerang effect \cite{hwang_solitary_2018}. 

Figure~\ref{fig:multifunc}a shows the changes in the experimental setup for energy harvesting. A magnet and coil enables us to harvest energy from the impact through electromagnetic induction. The voltage is measured across a 100 $\Omega$ resistor connected to the ends of the coil. Stiffening gradation in the onsite potential is implemented by adjusting the angle of the supporting rails until transition waves reflect back, as illustrated in Fig.~\ref{fig:multifunc}a. 

Figure~\ref{fig:multifunc}b shows the tip velocity of the multistable (with precompression) and linear (without precompression) metamaterials. The tip velocity in the multistable metamaterial under boomerang effect is 45\% lower than the corresponding linear metamaterial. This reduction can be attributed to the energy being dissipated more gradually  through phonon radiation in three stages: when the transition wave is generated, when the wave reverses direction, and when the metamaterial resets (supported by the model in Sec.~IV of the Supplemental Material). 

The high snap-through speed of the bistable units increases the rate of change of magnetic flux through the coil. As a result, the peak voltage of the energy harvested in the multistable metamaterial is 2.5 times that obtained from the corresponding linear metamaterial (Fig.~\ref{fig:multifunc}c). Thus, the multistable metamaterial outperforms the linear metamaterial in both shock absorption and energy harvesting.

\section{Conclusions}\label{sec:conc}
In this work, we leveraged the dynamics of transition waves to improve the damping performance and energy harvesting capabilities of multistable metamaterials subjected to impulsive excitations. We identified energy trapping and energy splitting as the two phenomena that enable the multistable metamaterial to damp shock response better than the corresponding linear metamaterial. Using analytical, numerical, and experimental methods, we showed that the judicious design of soft and stiff defect regions allows for trapping of transition waves at prescribed locations in the metamaterial. Furthermore, numerical simulations revealed that energy splitting occurs when multiple impulses or a single large-amplitude impulse acts on the metamaterial, resulting in the generation of fast-dissipating breathers. Interestingly, we observed that energy splitting can limit the metamaterial's response amplitude, regardless of the amplitude of the impulsive force. This offers a new pathway for achieving extreme damping in structures. In addition to damping, the rapid snap-through of the bistable units during transition wave propagation enables superior energy harvesting from impulsive excitations in the multistable metamaterial compared to its linear counterpart. Our experiments demonstrated a multifunctional metamaterial that automatically resets after every impulse by redirecting transition waves. Overall, these results establish that controlling transition wave dynamics provides a powerful framework for designing multifunctional metamaterials capable of energy trapping, extreme damping, and input-independent energy harvesting. 




\begin{acknowledgments}
The authors gratefully acknowledge the support from Purdue University's Frederick N. Andrews Fellowship and NSF CAREER award No. 1944597. 
\end{acknowledgments}


\bibliography{apssamp}

\end{document}


\preprint{APS/123-QED}

\title{\textbf{Transition Waves for Energy Trapping and Harvesting -- Supplemental Material} 
}

\author{Sneha Srikanth}
\author{Andres F. Arrieta}%
 \email{Contact author: aarrieta@purdue.edu}
\affiliation{%
 School of Mechanical Engineering, Purdue University, West Lafayette, Indiana 47907, USA.
}%

\date{\today}

\maketitle

\section{Generation of transition wave under impulsive forcing}\label{app:gen}
\begin{figure*}[b]
    \centering
    \includegraphics[width=\linewidth]{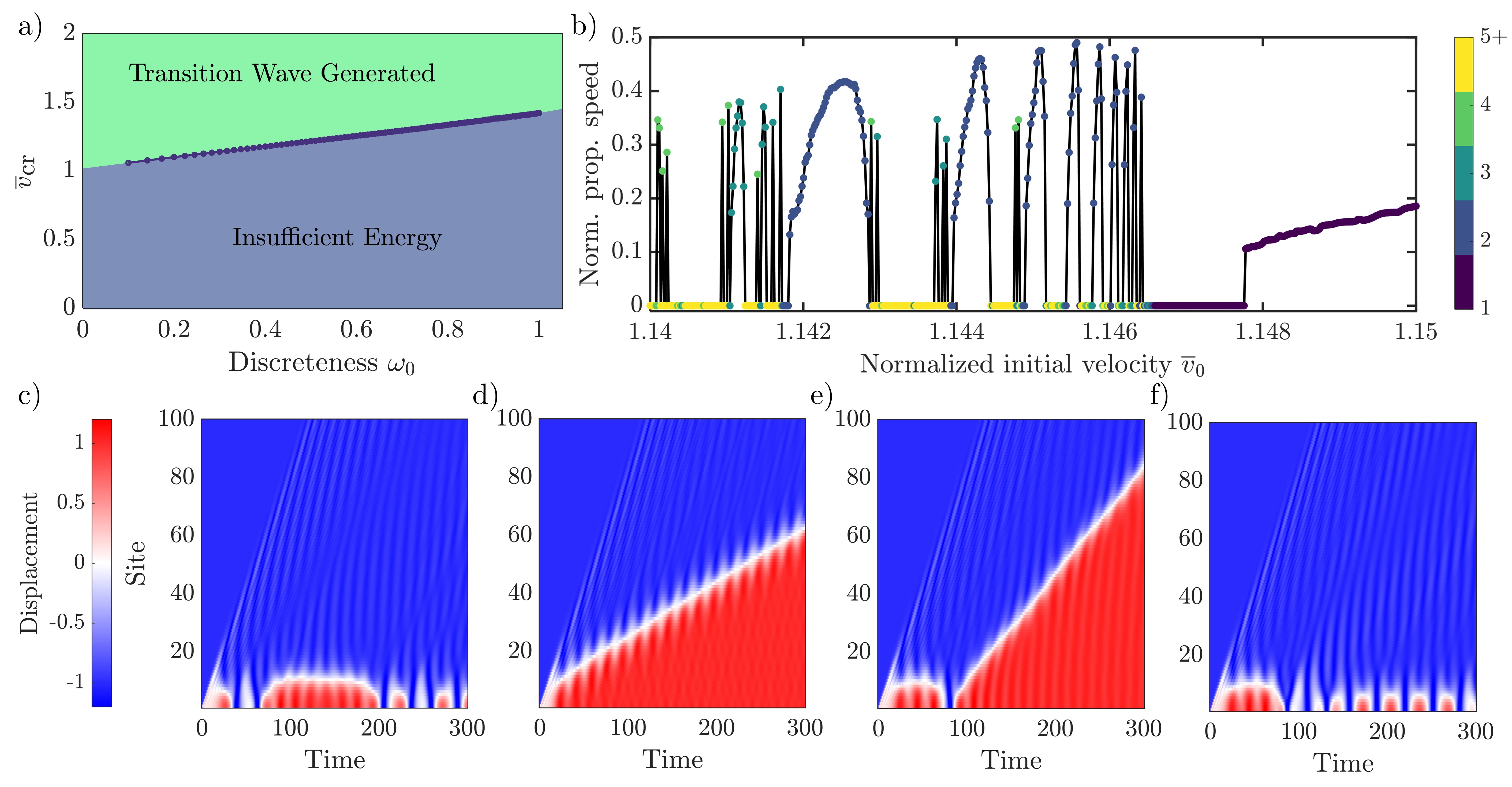}
    \caption{a) Relation between the critical value of the normalized initial velocity required to generate a transition wave and the discreteness of the multistable metamaterial. b) Fractal basin boundary demarcating the onset of transition wave generation on fine variation of the normalized initial velocity $\bar v_0$. Colors indicate the number of bounces. c) Boundary-oscillon for $\bar v_0=1.14$ and d) propagation of transition wave for $\bar v_0=1.15$. e) Transition wave in a 2-bounce window ($\bar v_0=1.145$)  and f) boundary-oscillon at $\bar v_0=1.1452$. In (b)-(f), $\omega_0^2 = 0.1$.}
    \label{fig:fractal} 
\end{figure*}

To study the generation of transition waves, we examine a uniform metamaterial with different values of discreteness subjected to different impulse strengths. Energy trapping can occur only when the impulse (or any external force) has sufficient energy to generate a transition wave. Such an impulse corresponds to a critical value $\bar v_\textrm{cr}$ of normalized initial velocity of the first unit. We observe that the energy of the transition wave (in Eq.~(9) of the main manuscript) and therefore the value of $\bar v_\textrm{cr}$ (shown in Fig.~\ref{fig:fractal}a) increases with $\omega_0$. Transition wave generation and therefore trapping are possible for $\bar v_0\ge\bar v_\textrm{cr}$. 

Unlike in the continuous $\phi^4$ model, transition waves or kinks in the discrete $\phi^4$ model can lose energy during their propagation by emitting phonons due to their interaction with the Peierls-Nabarro potential \cite{kivshar_peierls-nabarro_1993}. In particular, when the metamaterial is highly discrete ($\omega_0\gg 1$), this interaction can cause the transition wave to reverse, or be pinned. 
Thus, we restrict to the regime $\omega_0\in [0,1]$ for reliable generation and trapping of transition waves. 

When varying the initial velocity, or any other parameter, the boundary demarcating the onset of transition wave generation is fractal in nature in the undamped multistable metamaterial (Fig.~\ref{fig:fractal}b). Similar fractal structures have been reported in studies on kink-antikink collisions \cite{goodman2005kink} and kink-boundary interaction \cite{dorey_boundary_2017} in the continuous $\phi^4$ system. Figure~\ref{fig:fractal}b tracks the normalized speed of propagation of the transition wave and also the number of `bounces', i.e., the number of times the transition wave escapes the boundary, as a function of the normalized initial velocity. Insufficient energy in the transition wave can lead to a bound state called the boundary-oscillon, shown in Fig.~\ref{fig:fractal}c. When $\bar v_0\ge\bar v_\textrm{cr}$, the transition wave propagates, escaping the boundary only once. For intermediate values of $\bar v_0$, we observe two-bounce windows (Fig.~\ref{fig:fractal}e). Each of these windows is surrounded by subsequently higher bounce windows in the fractal structure. Boundary-oscillons can exist between any of these $n$-bounce windows (an example is shown in Fig.~\ref{fig:fractal}f). Note that although $N=500$, Figs.~\ref{fig:fractal}c-\ref{fig:fractal}f show the behavior of just the first 100 units to clearly observe the bounces. We also note that including material damping in the metamaterial destroys the fractal nature of the basin boundary and makes it smooth. 


\newpage

\section{CAD model of the metamaterial}
\begin{figure}[h]
    \centering
    \includegraphics[width=0.5\linewidth]{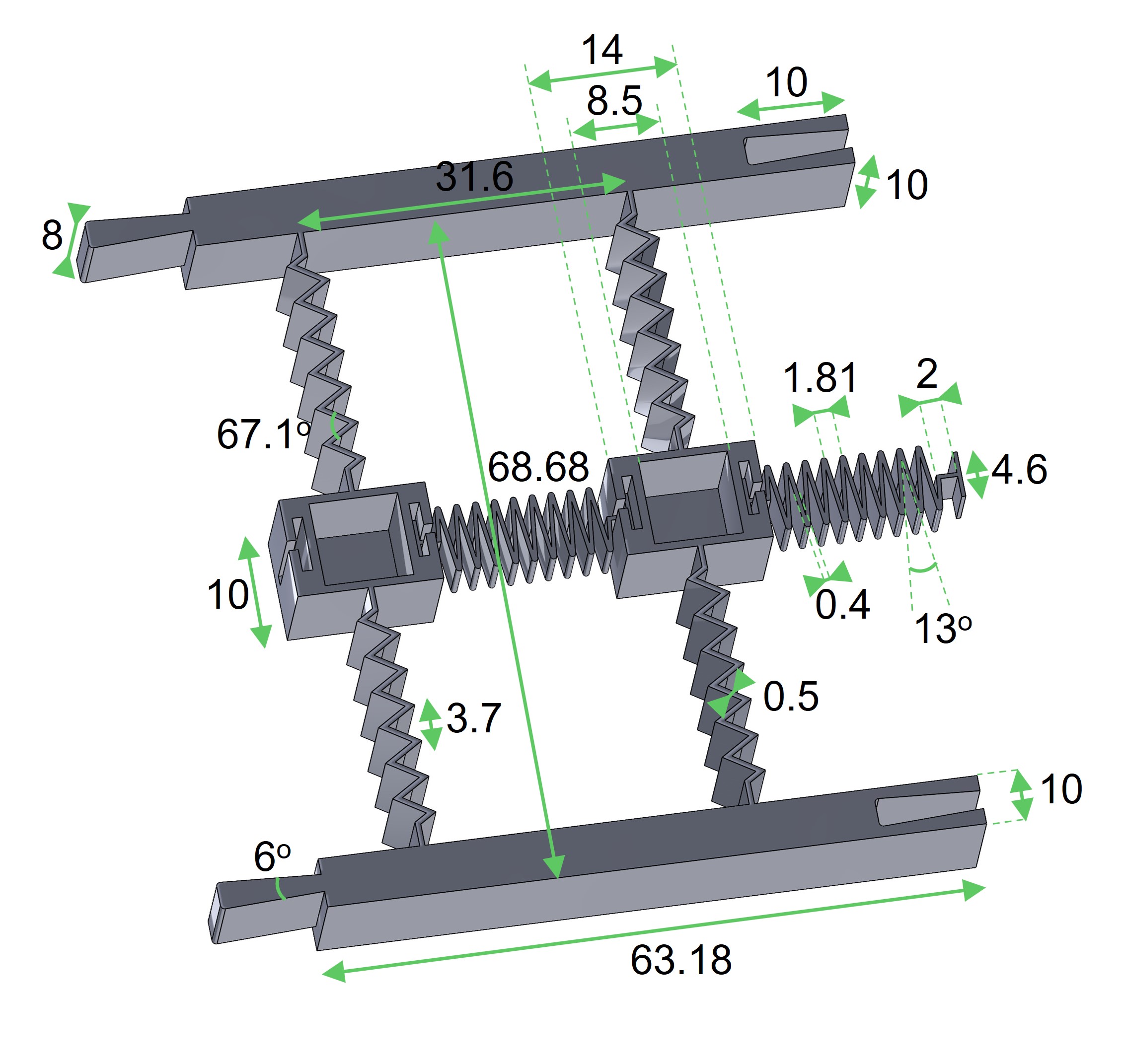}
    \caption{CAD model of two units and intersite spring showing key geometric dimensions}
    \label{fig:cad} 
\end{figure}

Figure~\ref{fig:cad} shows the CAD model and geometric dimensions of a chain of two units connected by intersite springs. Each unit's center has a cubical slot for inserting the mass. Intersite springs are inserted into the thin slots on consecutive units. Dovetail joints at the ends connect several chains to form a metamaterial of required number of units. 

\newpage

\section{Estimation of stiffness of intersite springs}

\begin{figure*}[h!]
    \centering
    \includegraphics[width=\linewidth]{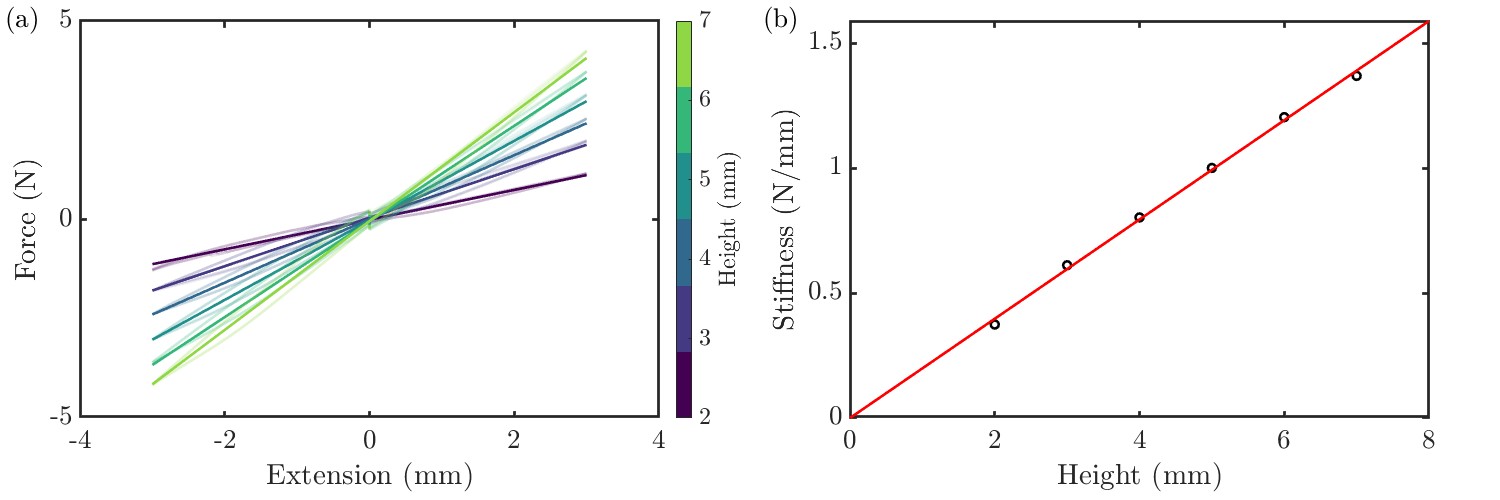}
    \caption{a) Force-extension curves (light colors) overlaid with fitted lines (dark colors) for intersite springs of different heights. b) Stiffness of the intersite spring as a function of its height. The red line represents the linear fit to the data.}
    \label{fig:instron} 
\end{figure*}

The Instron axial testing machine was used to estimate the stiffness of intersite springs with different heights. Each spring was subjected to three cycles of extension and compression at a rate of 5 mm/min, with 2 minutes of rest each time the extension returned to zero. Figure~\ref{fig:instron}a shows the force–extension curves and their fitted linear relationships for spring heights ranging from 2 mm to 7 mm. The slopes of the fitted lines reveal the intersite stiffness. Figure~\ref{fig:instron}b plots the stiffness of the intersite spring as a function of height. The stiffness increases linearly with spring height.

\newpage
\section{Boomerang effect }

\begin{figure*}[h!]
    \centering
    \includegraphics[width=\linewidth]{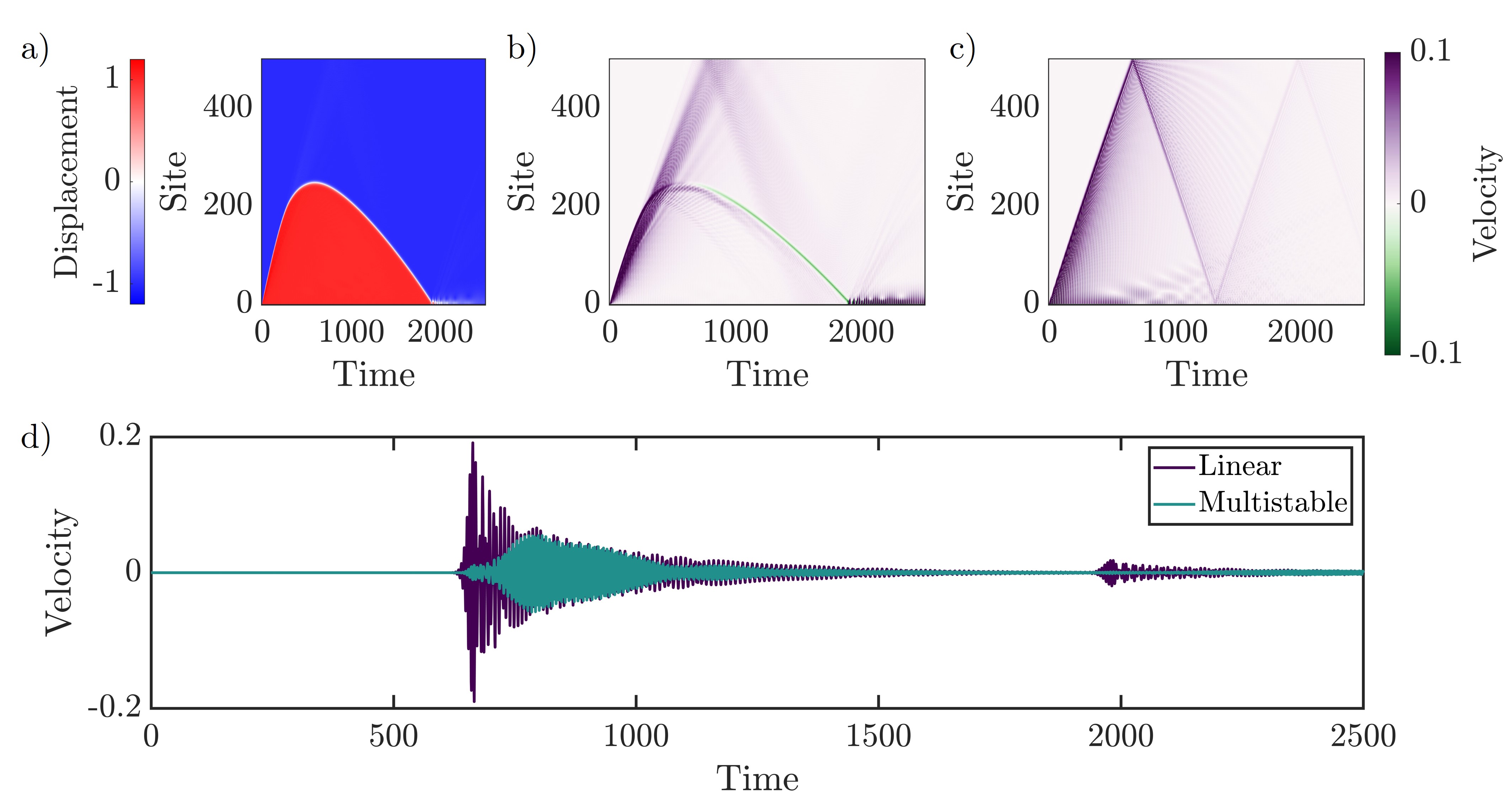}
    \caption{a) Boomerang effect in multistable metamaterial. b) Velocity of the units in the multistable metamaterial undergoing boomerang effect. c) Velocity of the units in the corresponding linear metamaterial. d) Comparison of tip velocities in the multistable and linear metamaterials. $N=500$, $\overline{b}=0.003$. Stiffening gradation of onsite bistable potential is applied by linearly varying $\omega_0^2$ from 0.1 to 0.2.} 
    \label{fig:boomerang} 
\end{figure*}

Multistable metamaterials with stiffening gradation exhibit boomerang effect, wherein transition waves slow down and reverse their direction in lieu of propagating forward (Fig.~\ref{fig:boomerang}a). Phonons are emitted gradually as the transition wave reverses direction and when the metamaterial resets (visible in Fig.~\ref{fig:boomerang}b). On the other hand, the corresponding linear metamaterial (Fig.~\ref{fig:boomerang}c) transmits the energy as a single stream of phonons generated at the time of impulse (at $t=0$). As a result, the tip velocity of the multistable metamaterial is 68\% lower compared to the linear metamaterial (Fig.\ref{fig:boomerang}d). 

\bibliography{apssamp}